\documentclass[aps,prb,showpacs,twocolumn]{revtex4}
\usepackage{graphicx,color}
\usepackage{amssymb}
\usepackage{amsmath}

\begin{document}
\title{Landau level splitting due to graphene superlattices}
\author{G. Pal}
\author{W. Apel}
\author{L. Schweitzer}
\affiliation{Physikalisch-Technische Bundesanstalt (PTB),
Bundesallee 100, 38116 Braunschweig, Germany}
\begin{abstract}
The Landau level spectrum of graphene superlattices is studied using
a tight-binding approach. We consider non-interacting particles moving
on a hexagonal lattice with an additional one-dimensional
superlattice made up of periodic square potential barriers, which are
oriented along the zig-zag or along the arm-chair directions of
graphene. In the presence of a perpendicular magnetic field, such
systems can be described by a set of one-dimensional tight-binding
equations, the Harper equations. The qualitative behavior of the energy
spectrum with respect to the strength of the superlattice potential
depends on the relation between the superlattice period and the
magnetic length. 
When the potential barriers are oriented along the arm-chair direction
of graphene, we find for strong magnetic fields that the zeroth Landau
level of graphene splits into two well separated sublevels, if the width
of the barriers is smaller than the magnetic length. In this situation, 
which persists even in the presence of disorder, a plateau with zero
Hall conductivity can be observed around the Dirac point. This Landau
level splitting is a true lattice effect that cannot be obtained from
the generally used continuum Dirac-fermion model. 
\end{abstract}
\date{\today} \pacs{73.22.Pr Electronic structure of graphene, 81.05.ue
  Graphene, 73.21.Cd Superlattices, 73.43.Cd QHE-Theory and modeling} 
\maketitle

\section{Introduction}\label{sectionIntroduction}
Recently, the electronic and transport properties of graphene
superlattices have been the subject of intense investigation. 
Theoretically, it was shown that the presence of periodic 
electrostatic~\cite{BaiZhangPRB2007,LouieSuperlattNature2008,
ParkLouiePRL2008,BarbierPeetersPereiraPRB2008,BreyFertigPRL2009,
LouieSuperlatB_PRL2009,HoChiuTsaiLinPRB2009,BarbierPeetersPRS2010,
BarbeierVasilopoulosPeetersPRB2010,WangZhuPRB2010,SunFertigBreyPRL2010,
GuineaLowPTRSA2010,BursetBreyFertigPRB2011,GuoLiuLiAPL2011} or
vector~\cite{MasirMatulisPeetersPRB2008,GhoshSharmaJPhysCondMatt2009,
deMartinoPRB2009,SnymanPRB2009,TanParkLouiePRB2010,DeMartinoPRB2011}
potentials, and also of periodic arrays of 
corrugations~\cite{IsacssonJonsonPRB2008,GuineaVozmedianoPRB2008,
BreyPalaciosPRB2008,TitovPRB2010,WangDevelPRB2011} tailors the
graphene properties in a unique way, leading to novel features and
interesting physics. In one-dimensional superlattices, i.e.,
two-dimensional (2D) superlattice potentials depending on only one
spatial direction, the Dirac cones of graphene are distorted, and
hence the velocity of a particle moving parallel to the potential
steps is reduced.~\cite{LouieSuperlattNature2008}  
Moreover, for certain superlattice parameters, this component of the
velocity is suppressed, and the carriers move only perpendicular to
the potential steps of the superlattice. Furthermore, for other
specific superlattice parameters, extra Dirac 
cones~\cite{BreyFertigPRL2009,HoChiuTsaiLinPRB2009,LouieSuperlatB_PRL2009}
and even Dirac lines~\cite{BarbierPeetersPRS2010} appear
in the energy spectrum of graphene besides the usual $K$ or $K'$ Dirac 
points that exist in the continuum model of graphene at the neutrality
point in the absence of a superlattice.

Interestingly enough, the emergence of the new Dirac points is
controlled by the ratio between the potential amplitude and the
superlattice period, irrespective of the superlattice profile, 
e.g., a cosine~\cite{BreyFertigPRL2009} or a 
Kronig-Penney~\cite{HoChiuTsaiLinPRB2009,LouieSuperlatB_PRL2009}
type, as long as the period is much larger than graphene's lattice
constant. The extra Dirac points and their associated zero-energy 
modes~\cite{BreyFertigPRL2009} drastically affect the transport 
properties~\cite{BreyFertigPRL2009,BarbeierVasilopoulosPeetersPRB2010,
BarbierPeetersPRS2010,WangZhuPRB2010} of the system and also the
Landau level sequence,~\cite{LouieSuperlatB_PRL2009} and hence the 
plateaus in the quantum Hall conductivity when a magnetic field is
applied. 

The implementation of two-dimensional superlattices is another route
to modulate the 
electronic properties of graphene. For example, for rectangular
superlattices, the velocity of carriers is also anisotropic and,
depending on the Fermi level, the charge carriers are electrons,
holes, or a mixture of both.~\cite{LouieSuperlattNature2008} 
Recently, it has been shown that for two-dimensional rectangular
superlattices the conductivity is unchanged from the result of
pristine graphene, even if the velocity renormalization induced by the
superlattice is quite large.~\cite{BursetBreyFertigPRB2011} 
Also, new Dirac points with and without energy gaps can emerge at
high-symmetry points in the Brillouin zone in two-dimensional 
triangular superlattices.~\cite{ParkLouiePRL2008,GuineaLowPTRSA2010}
Ab-initio studies of the electronic and magnetic properties of
graphene-graphane superlattices have also been
reported.~\cite{NievesPartoensPeetersPRB2010,LeeGrossmanAPL2010} 
For instance, it was shown that the zig-zag or arm-chair orientation
of the graphene-graphane interface has a significant impact on the
electronic properties of the system.\cite{LG11}

Experimentally, there are different possibilities to fabricate
graphene superlattices. For example, it is possible to imprint
superlattice patterns with periodicity as small as 5 nm using
electron-beam induced hydrocarbon lithography on graphene
membranes.~\cite{Experment2APL2008}  Graphene grown epitaxially on 
Ru(0001)~\cite{Experment3PRB2007,Experiment1PRL2008,
Experment4NatMat2008,Experiment5PRL2008} or 
Ir(111)~\cite{Experment6NanoLett2008,Experment7NewJPhys2008,
Experment8PRL2009} surfaces, and also on SiC~\cite{deHeerNature2007,
Experiment9PRB2009} shows two-dimensional superlattice patterns
with lattice period of a few nanometers and potential strength in the
range of few tenths of an electron volt. In suspended graphene, the
existence of periodic ripples has been recently
demonstrated.~\cite{BaoNatureNanotech2009} 
Another possibility to make superlattices with controlled potential
amplitude is to fabricate periodically patterned gates: 
\emph{p-n} and \emph{p-n-p} junctions in 
graphene~\cite{MarcusScience2007,PhysRevLett.98.236803,
YoungKimNatPhys2009,PhysRevLett.102.026807,TaruchaNJPhys2009}
have already been realized.

In the present work, we study the evolution of the Landau levels of
graphene appearing in the presence of a one-dimensional superlattice
and a strong magnetic field applied perpendicular to the graphene
plane. For the superlattice, we assume a Kronig-Penney type 
of electrostatic potential with alternating barriers of $+V$ and $-V$
potential strengths and barrier width $L_a$. Here, two cases are
considered, one with the potential barriers oriented along the zig-zag
(zz) and the other oriented along the arm-chair (ac) directions of
graphene. We use a tight-binding approach with nearest-neighbor
hopping. The magnetic field is incorporated in the hopping integral
through the Peierls phases as usual. Since the superlattice depends 
only on one direction, we use plane waves in the other
direction. Thus, the two-dimensional problem is reduced to a set of
one-dimensional equations, known as the Harper
equations.\cite{HarperPaper} 

There are two characteristic lengths in the system: one is the
superlattice barrier width $L_a$ and the other is the magnetic length
$l_B$. The behavior of the energy spectrum is governed by the
relation between $L_a$ and $l_B$. We find that if $L_a$ is greater
than $l_B$, the 
Landau levels acquire a finite broadening (irrespective of disorder)
when the superlattice potential strength increases. In this case, the
Landau level sequence disappears already for small values of $V$. In
the other case, when $L_a$ is smaller than $l_B$, the Landau levels
picture survives for much higher values of the potential strength.
Also, two or more Landau levels may merge together when $V$ is
increased. Consequently, the plateaus in the Hall conductivity show an
unconventional sequence that may be controlled by either the external
electrostatic potential or the applied magnetic field. When the
barriers of the superlattice potential are along the arm-chair
direction of graphene, and when $L_a<l_B$, we find that the zeroth
Landau level splits into two sublevels, a novel signature of the
interplay between the magnetic field and superlattice potential
orientation. We show that the splitting is robust against different
types of disorder affecting the superlattice potential. This holds
also for a two-dimensional chess-board type superlattice.

The paper is organized as follows. In Section~\ref{sectionTheory}
we introduce the theoretical model that we use in order to investigate
graphene superlattices in the presence of a strong magnetic field. In 
Section~\ref{sectionResults} we present the results concerning the
energy spectrum of the system for various superlattice parameters and
magnetic flux density strengths. A summary and concluding remarks are
given in Section~\ref{sectionSummary}.

\section{Harper equation for graphene ribbons}\label{sectionTheory}
The Harper equations of graphene ribbons in a uniform magnetic field
reduce the two-dimensional tight-binding problem to two 
one-dimensional equations due to the translational invariance along
one axis. Derivations of the Harper equations can be found in
Ref.~\onlinecite{Rammal} for a hexagonal lattice with zig-zag edges
and in Ref.~\onlinecite{Harper1Japans} for bricklayer lattices with
zig-zag and with arm-chair edges. Here, we briefly derive Harper
equations in the presence of a superlattice potential. From these,
we calculate the electronic energy spectra of monolayer graphene
ribbons with oriented edges in a perpendicular magnetic field and a
superlattice potential.  

\begin{figure}
\includegraphics[width=1.\columnwidth]{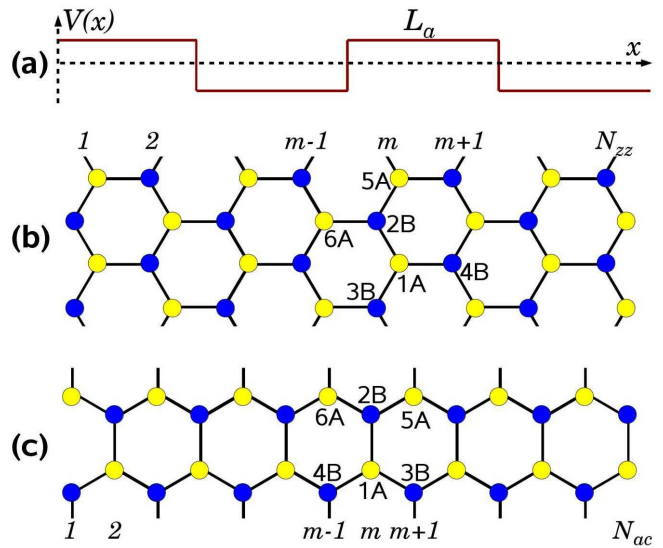}
\caption{(a) The profile of the Kronig-Penney superlattice potential. 
(b) Geometry of the zz ribbon with $N_{\rm zz}$ zz lines. 
(c) Geometry of the ac ribbon with $N_{\rm ac}$ dimer lines.
Periodic boundary conditions are applied in the $y$ direction with 
length $L$ generally large compared to the width $W$ of the sample.
\label{FIGgeomZZAC}}
\end{figure}

We consider a one-band tight-binding model with nearest neighbor
hopping on a hexagonal lattice. The structure of graphene ribbons 
of length $L$ with zig-zag (zz) and arm-chair (ac) edges is shown in 
Fig.~\ref{FIGgeomZZAC}. The two inter-penetrating triangular
sublattices are denoted by A and B. The ribbon width $W\ll L$ ($x$
direction) is defined by the number $N_{\rm zz}$ of zz lines for the
zz ribbon and by the number $N_{\rm ac}$ of dimer lines for the ac
ribbon   
\begin{align}
 W_{\rm zz}=& \big( \frac{3}{2}N_{\rm zz}-1\big) a\\
 W_{\rm ac}=& (N_{\rm ac}-1)\frac{\sqrt{3}}{2}a,
\end{align}
where $a$ is the distance between two neighboring carbon atoms
($a = 1.41$\,\AA{}). 

With $\psi({\bf r}_i)$ the wavefunction amplitude on site ${\bf r}_i$,
the Schr\"{o}dinger equation reads
\begin{equation}\label{SchEq}
 \varepsilon \psi({\bf r}_i)=V({\bf r}_i) \psi({\bf r}_i) + \sum_{j}
 t_{ij} \psi({\bf r}_j), 
\end{equation}
where $\varepsilon$ is the associated eigenvalue and $V$ the
electrostatic potential (superlattice). The sum runs over the nearest
neighbors of atom $i$ and $t_{ij}$ is the transfer integral between
atoms $i$ and $j$. The magnetic field perpendicular to the graphene
plane is incorporated in the hopping term by means of the Peierls
phase 
\begin{equation}\label{Peierls_1}
 t_{ij}\rightarrow t e^{i\gamma_{ij}},
\end{equation}
where $t$ is the hopping parameter ($t=$2.7\,eV) and $\gamma_{ij}$ is
given by the line integral of the vector potential from site ${\bf
  r}_i$ to site ${\bf r}_j$, and $\phi_0=h/e$ is the magnetic flux
quantum  
\begin{equation}\label{Peierls_2}
 \gamma_{ij}=\frac{2 \pi}{\phi_0}\int_{{\bf r}_i}^{{\bf r}_j} {\bf A} \cdot d{\bf l}.
\end{equation}
We use the  Landau gauge $ {\bf A}=(0,Bx,0)$ and the
translational invariant direction of each ribbon is taken as the
$y$ axis. With this particular choice of the gauge, the line integral
of the vector potential between two nearest neighbors $i$ and $j$
becomes 
\begin{equation}\label{Peierls_3}
 \gamma_{ij}=\frac{2 \pi}{\phi_0}B(y_j-y_i)\frac{x_i+x_j}{2}.
\end{equation}
The magnetic flux $\phi=BS$ through the area $S=3a^2\sqrt{3}/2$ of a 
hexagon is taken to be a rational multiple of the flux quantum
$\phi=(p/q)\phi_0$, hence the magnetic flux density $B$ and the
magnetic length are set by the integers $p$ and $q$ which are chosen  
to be mutually prime: 
\begin{align}\label{BlB}
 B=& \frac{p}{q}\frac{h}{e} \frac{2}{3a^2\sqrt{3}} \\
 l_B=& \sqrt{\frac{\hbar}{eB}}=\sqrt{\frac{3a^2\sqrt{3}q}{4\pi p}}. \nonumber
\end{align}

The one-dimensional superlattice potential is taken to be periodic
along the $x$ direction, with periodicity $2L_a$ (barrier width $L_a$)
and constant in the $y$ direction  
\begin{equation}
 V(x_i)=V(x_i+2L_a).
\end{equation}
Figure~\ref{FIGgeomZZAC}~(a) shows the profile of such a superlattice
formed by a Kronig-Penney type of electrostatic potential. 
Because the Hamiltonian does not depend on $y$, we use plane waves for
the wavefunctions in the $y$ direction 
\begin{equation}\label{eiky}
 \psi({\bf r})= \psi(x) e^{ik_y y}.
\end{equation}
We are ready now to write the Harper equations for the zz and the
ac ribbon.

\subsection{Harper equations for the zz ribbon}
In the following, we label the atoms as shown in
Fig.~\ref{FIGgeomZZAC}~(b) and use $m$ to index the zz chains, where
$m$ takes values between $1$ and $N_{\rm zz}$. Then, the
Schr\"{o}dinger equations for the atoms 1A and 2B are 
\begin{align}\label{Heqzz1}
 & \varepsilon \psi_1^A(x_m^A)= V(x_m^A) \psi_1^A(x_m^A) + 
e^{i\gamma_{1,2}} e^{i k_y \frac{\sqrt{3}}{2}a} \psi_2^B(x_m^B)  \nonumber \\
 & + e^{i\gamma_{1,3}} e^{-i k_y \frac{\sqrt{3}}{2}a} \psi_3^B(x_m^B)
+ e^{i\gamma_{1,4}} \psi_4^B(x_{m+1}^B)\\
 & \varepsilon \psi_2^B(x_m^B)= V(x_m^B) \psi_2^B(x_m^B) + 
e^{i\gamma_{2,1}} e^{-i k_y \frac{\sqrt{3}}{2}a} \psi_1^A(x_m^A)  \nonumber \\
 & + e^{i\gamma_{2,5}} e^{i k_y \frac{\sqrt{3}}{2}a} \psi_5^A(x_m^A) 
+ e^{i\gamma_{2,6}} \psi_6^A(x_{m-1}^A). \nonumber
\end{align}
From Eq.~\eqref{Peierls_3}, the Peierls phases are
\begin{align}\label{Peierls_zz}
& \gamma_{1A,2B}=\gamma_{2B,5A}=2\pi \frac{p}{q}\frac{1}{3a}\frac{x_m^A+x_m^B}{2} \nonumber \\
& \gamma_{2B,1A}=\gamma_{1A,3B}=-2\pi \frac{p}{q}\frac{1}{3a}\frac{x_m^A+x_m^B}{2}  \\
& \gamma_{1A,4B}=\gamma_{2B,6A}=0. \nonumber
\end{align}
We denote $\psi_1^A(x_m^A)=\psi_m^A$, $\psi_2^B(x_m^{B})=\psi_m^B$
and so the Harper equations take the simple form 
\begin{align}\label{Heqzz2}
 & \varepsilon \psi_m^A = V(x_m^A) \psi_m^A +a_m \psi_m^B +\psi_{m+1}^B  \\
 & \varepsilon \psi_m^B = V(x_m^B) \psi_m^B +a_m \psi_m^A +\psi_{m-1}^A, \nonumber 
\end{align}
where we defined
\begin{equation}\label{am_zz}
a_m=2\cos \big[ k_y\frac{\sqrt{3}a}{2}+2\pi
  \frac{p}{q}\frac{1}{3a}\frac{x_m^A+x_m^B}{2} \big]. 
\end{equation}
Here, we use $x_m^A=(3m/2-1)a$ and $x_m^B=(3/2)(m-1)a$.

Equation~\eqref{Heqzz2} does not contain any boundary conditions
yet. In the case of zz ribbons with finite $N_{\rm zz}$, Dirichlet boundary
conditions are imposed by setting $\psi_{N_{\rm zz}+1}^{A/B}=\psi_0^{A/B}=0$, 
since there are no atoms for $m=0$ and $m=N_{\rm zz}+1$.
Including this fact, Eq.~\eqref{Heqzz2} simplifies for $m=1$
and $m=N_{\rm zz}$ to 
\begin{align}\label{dbc_zz}
 & \varepsilon \psi_{N_{\rm zz}}^A = V(x_{N_{\rm zz}}^A) \psi_{N_{\rm
      zz}}^A +a_{N_{\rm zz}} \psi_{N_{\rm zz}}^B  \\ 
 & \varepsilon \psi_1^B = V(x_1^B) \psi_1^B +a_1 \psi_1^A. \nonumber
\end{align}

For periodic boundary conditions (torus), one sets 
$ \psi_{N_{\rm zz}+1}^{A/B}=\psi_1^{A/B}$ and $\psi_0^{A/B} = \psi_{N_{\rm zz}}^{A/B}$
and Eq.~\eqref{Heqzz2} becomes
\begin{align}\label{pbc_zz}
 & \varepsilon \psi_{N_{\rm zz}}^A = V(x_{N_{\rm zz}}^A) \psi_{N_{\rm
      zz}}^A +a_{N_{\rm zz}} \psi_{N_{\rm zz}}^B +\psi_1^B \\ 
 & \varepsilon \psi_1^B = V(x_1^B) \psi_1^B +a_1 \psi_1^A +\psi_{N_{\rm zz}}^A . \nonumber
\end{align}
However, a more careful treatment is needed at the edge sites for
periodic boundary conditions, due to the requirement for
commensurability between the magnetic and the lattice lengths: the
Peierls phases must be periodic with the ribbon width. 
This implies that
\begin{equation}\label{comensurability_zz1}
e^{i\gamma_{mA,mB}}=e^{i\gamma_{(m+N_{\rm zz})A,(m+N_{\rm zz})B}},
\end{equation}
and one finds the condition
\begin{equation}\label{comensurability_zz2}
N_{\rm zz}=2 n \frac{q}{p},
\end{equation}
where $n$ can be any nonzero positive integer. This means that the
parameters for the magnetic flux density ($p/q$) and the ribbon
width (given by $N_{\rm zz}$) are not independent but must be chosen in
such a way that Eq.~\eqref{comensurability_zz2} holds. In order to
obtain the energy spectrum of the system, we diagonalize numerically
the RHS of Eq.~\eqref{Heqzz2}, i.e., a $2N_{\rm zz}\times 2N_{\rm zz}$
matrix with the appropriate boundary conditions.

\subsection{Harper equations for the ac ribbon}
To obtain the Harper equations for ac ribbons, we proceed in a similar
way as in the case of zz edges. Here, $m$ indexes now the dimer
lines of the ac ribbon, and takes values between $1$ and $N_{\rm ac}$. 
The Schr\"{o}dinger equations for the atoms 1A and 2B of
Fig.~\ref{FIGgeomZZAC}~(c) are ($x_m^A=x_m^B=x_m$): 
\begin{align}\label{Heqac1}
& \varepsilon \psi_1^A(x_m)= V(x_m) \psi_1^A(x_m) + e^{i\gamma_{1,2}} e^{i k_y a} \psi_2^B(x_m)  \\
& +   e^{i\gamma_{1,3}} e^{-i k_y \frac{a}{2}} \psi_3^B(x_{m+1})+
  e^{i\gamma_{1,4}} e^{-i k_y \frac{a}{2}} \psi_4^B(x_{m-1})\nonumber  \\
& \varepsilon \psi_2^B(x_m)= V(x_m) \psi_2^B(x_m) + e^{i\gamma_{2,1}} 
e^{-i k_y a} \psi_1^A(x_m)  \nonumber   \\
& + e^{i\gamma_{2,5}} e^{i k_y \frac{a}{2}} \psi_5^A(x_{m+1})+ 
e^{i\gamma_{2,6}} e^{i k_y \frac{a}{2}} \psi_6^A(x_{m-1}) , \nonumber 
\end{align}
with the  Peierls phases
\begin{align}\label{Peierls_ac}
& \gamma_{1A,2B}=-\gamma_{2B,1A}=2\pi \frac{p}{q}\frac{2}{3\sqrt{3}a}x_m  \\
& \gamma_{2B,5A}=-\gamma_{1A,3B}=2\pi \frac{p}{q}\frac{1}{3\sqrt{3}a}\frac{x_m+x_{m+1}}{2} \nonumber \\
& \gamma_{2B,6A}=-\gamma_{1A,4B}=2\pi \frac{p}{q}\frac{1}{3\sqrt{3}a}\frac{x_m+x_{m-1}}{2} . \nonumber
\end{align}
Again, we denote $\psi_1^{A,B}(x_m)=\psi_m^{A,B}$. Then
the Harper equations take the simple form
\begin{align}\label{Heqac2}
& \varepsilon \psi_m^A = V(x_m) \psi_m^A +a_m \psi_m^B +b_m
  \psi_{m+1}^B+ b_{m-1} \psi_{m-1}^B  \\ 
& \varepsilon \psi_m^B = V(x_m) \psi_m^B +a_m^* \psi_m^A
  +b_m^*\psi_{m+1}^A+b^*_{m-1}\psi_{m-1}^A,  \nonumber  
\end{align}
where we defined
\begin{align}\label{am_ac}
a_m=&\exp \Big\{i\big[ k_y a +2\pi \frac{p}{q}\frac{2}{3a\sqrt{3}}x_m
  \big] \Big\} \\ 
b_m=&\exp \Big\{-i\big[ k_y\frac{a}{2} +2\pi \frac{p}{q}
  \frac{1}{3a\sqrt{3}} \frac{x_m+x_{m+1}}{2}\big] \Big\} ,\nonumber 
\end{align}
with $x_m=(\sqrt{3}/2)(m-1)a$.

For Dirichlet boundary conditions we use $\psi_0^{A/B}=0$ and
$\psi_{N_{\rm ac}+1}^{A/B}=0$, while for periodic boundary conditions
$\psi_0^{A/B}=\psi_{N_{\rm ac}}^{A/B}$ and
$\psi_{N_{\rm ac}+1}^{A/B}=\psi_{1}^{A/B}$. 
In the case of ac ribbons with periodic boundary conditions,
commensurability between the lattice and the magnetic field requires
that 
\begin{equation}\label{comensurability_ac1}
N_{\rm ac}=6 n \frac{q}{p}.
\end{equation}
The energy spectrum of an ac ribbon is obtained by diagonalizing the
RHS of Eq.~\eqref{Heqac2}, i.e., a $2N_{\rm ac}\times 2N_{\rm ac}$ matrix.

\section{Results and Discussions}\label{sectionResults}
In this section, we present results obtained by solving the Harper
equations corresponding to graphene superlattices under perpendicular
magnetic fields. The superlattice potential is given by a
Kronig-Penney function periodic along the $x$ direction, with barrier
width $L_a$, and with alternating $\pm V$ barrier heights. The
perpendicular magnetic field is set by the parameters $p$ and $q$,
according to Eq.~\eqref{BlB}. For simplicity, we take in the
following $p=1$. 

As a test, we consider first the cases with no superlattice potential
or no magnetic field. Then we show that, for $V\neq 0$ and $B\neq 0$,
the energy spectrum and hence the system properties strongly depend
on the relation between the superlattice barrier width $L_a$ and the
magnetic length $l_B$, as well as on the \textit{orientation} of the
superlattice barriers with respect to the zz or ac directions of
graphene.

\subsection{$V=0$ or $B=0$}
For infinite 2D systems, when the superlattice potential is not
present ($V=0$) and for high  magnetic fields, the energy eigenvalues
are the usual Landau level (LL) sequence which, according to the
continuum model, occur at\cite{HAL88,ZA02}
\begin{equation}\label{ELL}
 E^{LL}_n=\mathrm{sign}(n)\frac{\hbar v_F}{l_B}\sqrt{2|n|}
=\mathrm{sign}(n)\sqrt{\frac{2\pi \sqrt{3}p}{q}}\sqrt{|n|}t,
\end{equation}
where $n$ is the LL index, and 
for the last equality we have used Eq.~\eqref{BlB} and $\hbar v_F=3/2 t a$.

\begin{figure}
\includegraphics[width=0.8\columnwidth]{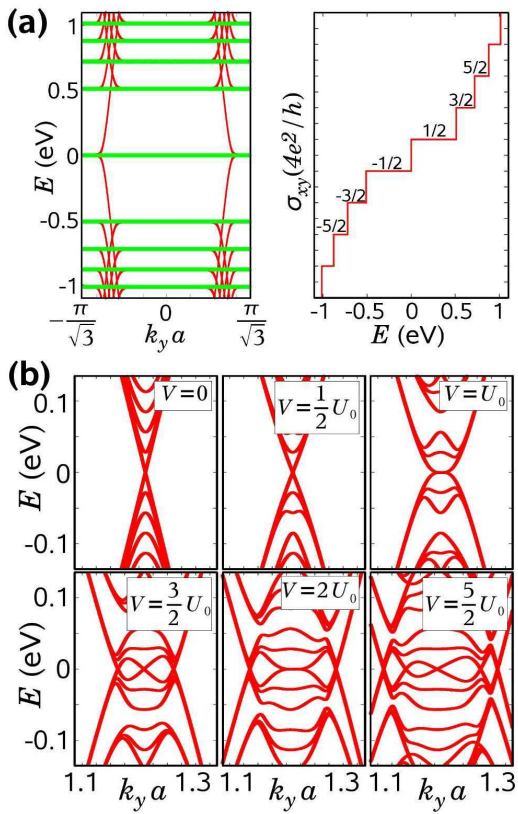}
\caption{(a) Left: the energy bandstructure of a graphene
  zig-zag ribbon of width $N_{\rm zz}=600$ in the presence of a
  strong magnetic field with $p/q=1/300$. The flat bands correspond
  to the bulk Landau levels. The energies of the $k_y$ dependent edge
  states are highlighted in red. Right: The corresponding Hall
  conductivity.   
  (b) The eigenvalues of a graphene Kronig-Penney superlattice with
  barrier width $L_a=12.7$\,nm (60 zz lines) for several barrier
  heights specified in each panel ($U_0=0.14$\,eV) in the absence of a
  magnetic field.  
\label{FIG_V0B0}}
\end{figure}

The left panel of Fig.~\ref{FIG_V0B0}~(a) shows the energy spectrum in
the $1^{\rm st}$ Brillouin zone of a graphene ribbon in the
absence of the superlattice potential. The low lying LL energies 
(shown in green, independent of $k_y$) appear according to
Eq.~\eqref{ELL}. Higher energies (not shown) may deviate from the
sequence because Eq.~\eqref{ELL} is valid for an infinite system. 
We treat here ribbons with a finite width and are only
interested in the low-energy domain. For Dirichlet boundary
conditions, the edge states ($k_y$ dependent) are shown in
red. Between two LLs, the Hall conductivity $\sigma_{xy}$ is
equal to $e^2/h\times$(number of edge states) and experiences a
jump every time the Fermi energy coincides with the LL energy. The
corresponding Hall conductivity is shown in the right panel of
Fig.~\ref{FIG_V0B0}~(a), with the quantized values of the
integer quantum Hall effect of graphene.~\cite{NovoselovQHE,NovoselovRMP}  

\begin{figure}[t]
\includegraphics[width=1.\columnwidth]{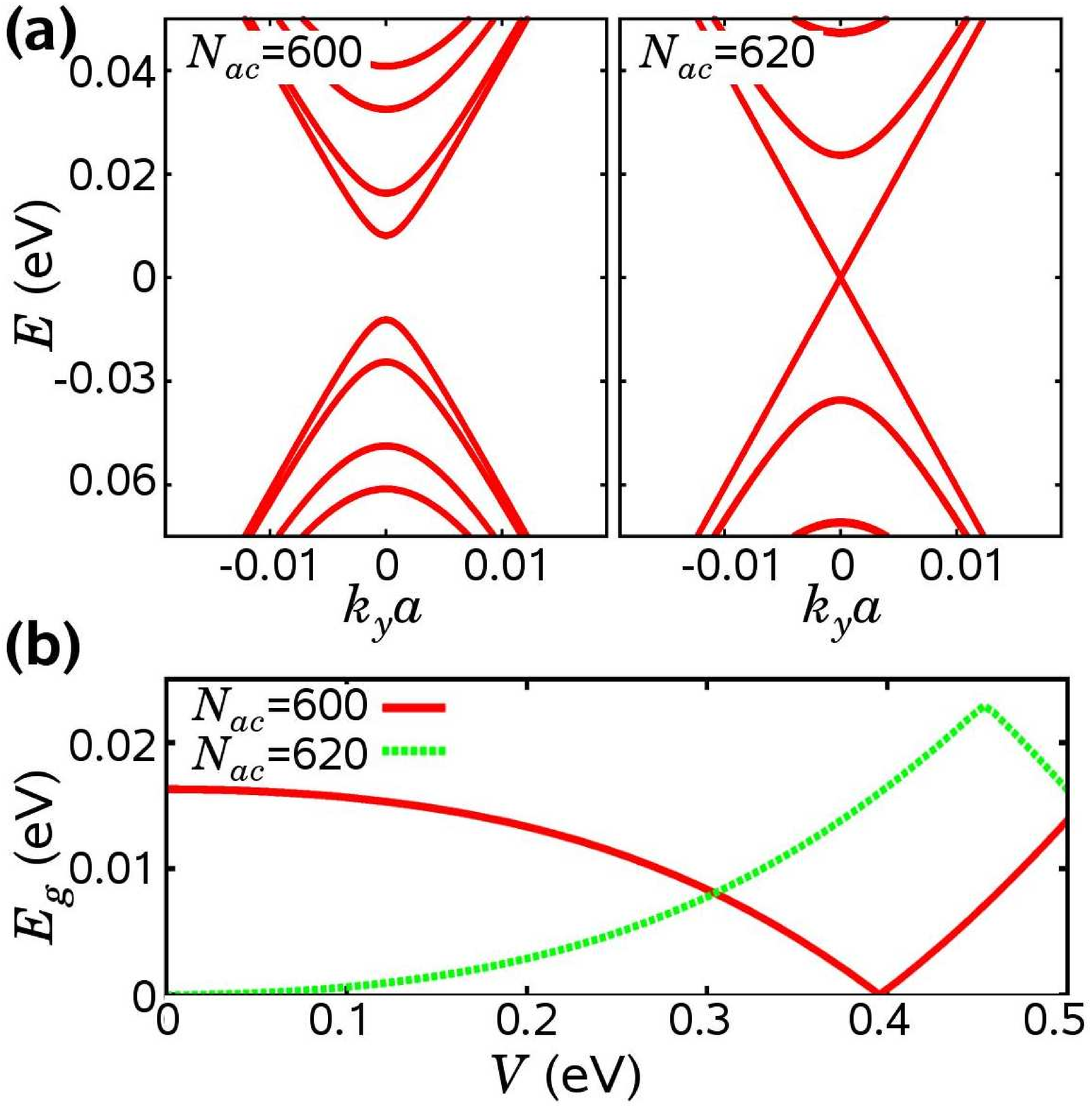}
\caption{(a) Energy bands of ac ribbons for $V=0$ with $N_{\rm
    ac}=600$ (semiconducting left) and $N_{\rm ac}=620$ (metallic
  right). (b) The energy gap $E_g$ for the ac ribbons from (a) in the
  presence of an additional superlattice with $L_a=1.22$\,nm as a
  function of the superlattice potential strength $V$.
\label{FIG_ACribbons}}
\end{figure}

In the absence of the magnetic field, and depending on the
superlattice parameters, the energy spectrum of a graphene zz
superlattice may exhibit additional Dirac points for $E=0$. For a
Kronig-Penney superlattice with barrier width $L_a$ and height $V$,
the number of Dirac points increases by two ($\times 4$ considering
valley and pseudospin degeneracy) whenever the potential amplitude
exceeds a value of 
\begin{equation}
 U_0=n \frac{3\pi t a}{2L_a}
\label{unull}
\end{equation}
with $n$ a positive integer.~\cite{LouieSuperlatB_PRL2009}
Figure~\ref{FIG_V0B0}~(b) illustrates this possibility of changing the 
number of Dirac points at $E=0$ by tuning the height of the
superlattice potential. Shown are the eigenvalues of the Harper
equation around the Dirac point for a zz ribbon with $N_{\rm zz}=600$
and for periodic boundary conditions in both directions. One
superlattice barrier contains 60 zz lines, which corresponds to a
barrier width of $L_a=12.7$\,nm. One clearly sees that when the
barrier height  $V$ exceeds integer multiples of $U_0=0.052\,t$
($=0.14$\,eV), then additional zero-energy modes appear in the energy
spectrum. Our spectra for $B=0$ are in agreement with the results
obtained by means of a continuum Dirac-equation approach for graphene
superlattices.\cite{LouieSuperlatB_PRL2009,BreyFertigPRL2009,
WangZhuPRB2010,BarbierPeetersPRS2010}

\subsection{$B=0$ and $V\neq 0$ ac ribbons}
The electronic properties of graphene nanoribbons with ac edges
strongly depend on their size. It is known that such systems are
metallic when $(N_{\rm ac}-2)/3\in \mathbb{Z}$ and semiconductor
otherwise.~\cite{BF06} Figure~\ref{FIG_ACribbons} (a) shows the energy
bands of two typical ac ribbons, a semiconducting one with
$N_{\rm ac}=600$ that corresponds to $W_{\rm ac}^{600}=73.6$ nm and an
energy gap $E_g=0.0162$~eV, and a metallic one with
$W_{\rm ac}^{620}=76.12$\,nm.  
A superlattice potential with $L_a=1.22$\,nm (10 ac dimer lines under
each barrier) with barriers along the ac edges is imposed on the
systems and the evolution of the energy gap around $E=0$ is calculated
as a function of the potential strength. The results are shown in
Fig.~\ref{FIG_ACribbons} (b). For the  semiconductor ac ribbon
($N_{\rm ac}=600$), the energy gap is reduced from $E_g=0.0162$\,eV
when $V=0$ to $E_g=0$ when $V=0.39$\,eV, and then increases again for
higher values of the potential strength.  For the metallic ac ribbon
($N_{\rm ac}=620$) the superlattice potential opens a spectral gap
that reaches a maximum value of $E_g=0.022$\,eV for $V=0.45$\,eV and
decreases for higher $V$. The respective minimal and maximal energy
gaps $E_g$ depend both on $N_{\rm ac}$ and $L_a$.

\subsection{Superlattice parallel to zig-zag edges}\label{subsection_ZZedges}
We now consider Kronig-Penney superlattices with potential barriers
oriented along the zz direction of graphene in a strong perpendicular
magnetic field. In our calculations we 
use graphene ribbons with $N_{\rm zz}=12000$, which corresponds to a
ribbon width of 2555\,nm. The Kronig-Penney superlattice parameters
are chosen in such a way that the number of barriers is even.
Also, the number of magnetic flux quanta per graphene plaquette is
fixed at $p/q=1/6000$, which gives $B=13.15$\,T for the magnetic field
and $l_B=7.07$\,nm for the magnetic length. These settings allow us to
study infinitely long (in the $y$ direction) ribbons with
Dirichlet or periodic boundary conditions in the transverse direction.

\begin{figure}[t]
\includegraphics[width=1.\columnwidth]{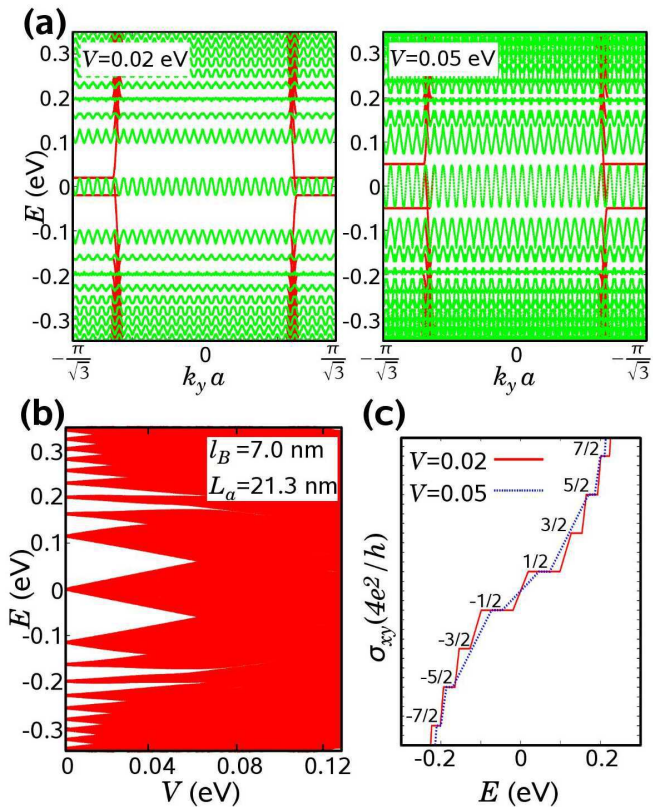}
\caption{(a) The energy bands of a graphene ribbon with superlattice
  with barrier widths of 21.3\,nm oriented along the zz edges. The
  applied magnetic field is $13.15$\,T and the superlattice potential
  strength is $0.02$\,eV (left) and $0.05$\,eV (right). The ribbon
  edge states are highlighted.  
  (b) The energy of the LLs as function of the superlattice potential
  strength. 
  (c) The energy-dependent Hall conductivity corresponding to the
  energy bands from (a).  
\label{FIG_ZZvsK}}
\end{figure}

First, we discuss the case when the barrier width is larger than the
magnetic length. Figure~\ref{FIG_ZZvsK} shows the results for
$L_a=21.3$\,nm (100 zz lines per barrier) and $l_B=7.07$\,nm. Typical
energy spectra as function of the wavevector $k_y$ in the first
Brillouin zone are given in 
Fig.~\ref{FIG_ZZvsK}~(a) for two different values of the potential
strength $V=0.02$\,eV (left) and $V=0.05$\,eV (right). The difference
between periodic and Dirichlet boundary conditions consists in the
appearance of edge states when applying the latter. The oscillations
seen in the LL energies correspond to mini-Brillouin zones 
imposed by the superlattice. The presence of the
superlattice modifies the energies of the LLs, which acquire now a
finite width depending on the value of $V$. Note that the finite
bandwidth of the LLs is a consequence of applying a superlattice and
not because of disorder, which is always present in experimental
situations and induces additional broadening of the LLs. The width of
the LLs increases with increasing $V$, and for larger values of the
potential strength the Landau bands merge together in the sense that
there remain no energy gaps between them.  For example, in
Fig.~\ref{FIG_ZZvsK} (a) and for $V=0.05$\,eV, this is the case for  
the $\pm 1^{\text{st}}$ and the $\pm 2^{\text{nd}}$ LLs. 
Interestingly enough, the broadening of the LLs is not the same for
all the LLs. For the parameters used in Fig.~\ref{FIG_ZZvsK} the width
of the $\pm 3^{\text{rd}}$ LL is smaller than that of the other 
LLs, and this particular LL will merge with the others only for higher
values of $V$. The origin of this individual broadening is discussed in
detail below.  

In Fig.~\ref{FIG_ZZvsK}~(b) we show the energies of the LLs as a
function of the superlattice potential strength. The energy gaps
between the LLs dissappear altogether in this case when $V\gtrsim
0.07$\,eV. Also, for $V<0.07$\,eV we find that the bandwidth of the
$0^{\text{th}}$ LL is equal to the value of $V$. 
When increasing the superlattice barrier width while $L_a\gg l_B$,
the energy bandwidth of the higher LLs also increases with the
strength of the potential, and the merging of the LLs occurs for
even smaller values of $V$.   

The unusual structure of the energy bands is most directly reflected
in the plateau sequence of the quantum Hall conductivity. 
Figure~\ref{FIG_ZZvsK} (c) schematically shows the Hall conductivity
versus the Fermi energy corresponding to the parameters used in panel
(a), i.e., $L_a=21.3$\,nm, $l_B=7.07$\,nm. Comparing the cases
$V=0.02$\,eV and $0.05$\,eV, one sees that with increasing $V$ the
reduction of the energy gaps between the LLs leads to a decrease of the
plateau widths. Moreover, for $V=0.05$\,eV the plateaus at $\pm (3/2)
(4e^2 /h)$ are not present, and the Hall conductivity has an
unconventional step size. This is because the  $\pm 1^{\text{st}}$ and
the $\pm 2^{\text{nd}}$ LLs are now merged together, and there is no
energy gap between them. Of course, the plateaus in the
energy-dependent Hall conductance must not be confused with the 
experimentally observed Hall plateaus, which originate from
localization due to the intrinsic disorder.

\subsubsection{Landau level broadening}
\begin{figure}[t]
\includegraphics[width=1.42\columnwidth,angle=270]{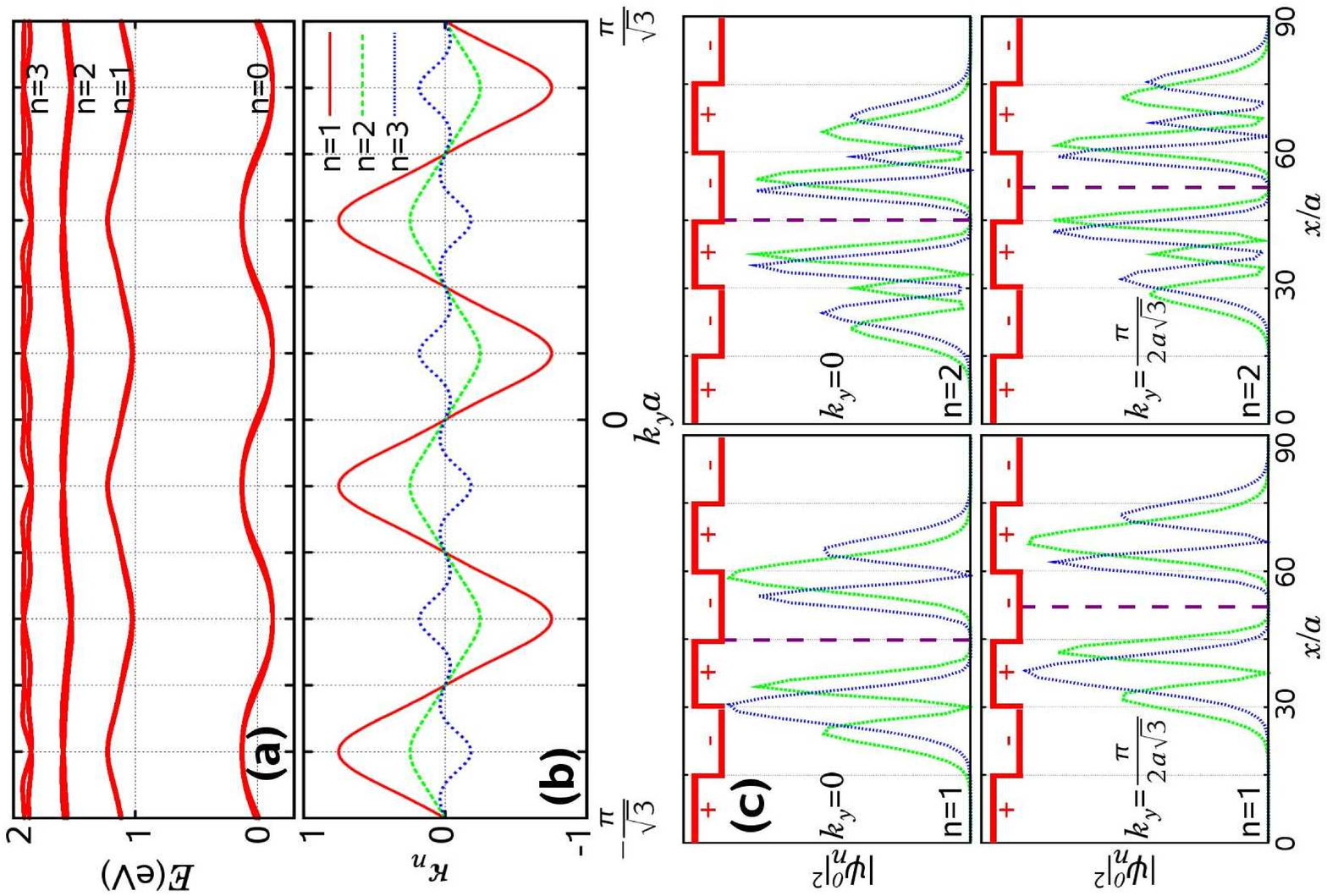}
\caption{(Color online) (a) The energies of the four lowest LLs of a
  system with superlattice potential $V=0.135$\,eV and $L_a=15$\,$a$. 
  The system size and magnetic field are $N_{zz}=120$ and $p/q=1/60$,
  respectively.   
  (b) The $\kappa$ integrals for the $1^{\rm st}$, $2^{\rm nd}$ and
  $3^{\rm rd}$ LLs calculated according to Eq.~\eqref{overlap}. 
  Here, $\kappa_1$ (red) has the largest amplitude and $\kappa_3$
  (blue) the smallest. 
  (c) The wavefunctions of the $1^{\rm st}$ (left) and the $2^{\rm
    nd}$ (right) LL as a function of the position $x$ for two values
  of $k_y$ where the corresponding $\kappa_n$ integrals are zero
  (upper part) and maximal (lower part). The contributions of
  the $A$ (blue) and the $B$ (green) sublattices to the wavefunction
  are separately shown. Please note that the lines connect only the
  amplitudes at the respective lattice sites. The superlattice
  potential is also sketched and the dashed lines are the reflection
  symmetry axes.  
\label{FIG_Broadening}}
\end{figure}

The different broadening of the distinct LLs labeled $n$ can be
explained using perturbation theory. We consider the superlattice
potential as a perturbation to the graphene wavefunctions $\psi^0_n$
belonging to a given energy. The integral  
\begin{equation}\label{overlap}
 \kappa_n(k_y)=\frac{1}{V} \int \psi^{0{\displaystyle *}}_n(x,k_y) V(x)
 \psi^0_n(x,k_y) dx  
\end{equation}
gives the energy corrections up to first order to one of the $n^{\rm
  th}$ LL energies as a function of $k_y$. The values of $|\kappa_n|$
for different LLs provide a quantitative measure of the influence of
the superlattice on the LL spectrum. The energies of the LLs with a
larger $|\kappa_n|$ are more spread out than the energy levels
corresponding to a smaller $|\kappa_n|$, as illustrated in
Fig.~\ref{FIG_Broadening}.  
Here, we consider a system with $N_{zz}=120$, $p/q=1/60$,
$V=0.135$\,eV and $L_a=15$\,a.  
The energies of the lowest four LLs given in Fig.~\ref{FIG_Broadening}
(a) show that the 0th LL ($E_0\sim 0$\,eV) is most broadened by the
superlattice, and the broadening decreases successively for the 
$1^{\rm st}$ LL ($E_1\sim 1.13$\,eV),  $2^{\rm nd}$ ($E_2\sim 1.56$\,eV) 
and $3^{\rm rd}$ ($E_3\sim 1.98$\,eV).

The amplitudes of $\kappa_n$ reflect directly the broadenings of
different LLs, as shown in Fig.~\ref{FIG_Broadening} (b). For example,
$\kappa_1$ has the largest amplitude in comparison with $\kappa_2$ and
$\kappa_3$, the latter has the smallest amplitude. Correspondingly, in
the presence of a superlattice potential, the $1^{\rm st}$ LL exhibits
a larger broadening than the $2^{\rm nd}$, and the $3^{\rm rd}$ LL
shows the smallest broadening. 

To understand the oscillations of the $\kappa_n$ integrals with
$k_y$, we analyze the spatial representation of the wavefunctions
$\psi^0_n$ of graphene in a perpendicular magnetic field. 
Figure~\ref{FIG_Broadening} (c) shows $|\psi^0_n(x)|^2$ for
$n=1$ (left) and $n=2$ (right) and for two values of $k_y$, which
correspond to zero (upper part) and maximum (lower part) values
of the respective $|\kappa_n|$ integrals. Note that changing $k_y$ only
shifts the wavefunctions with respect to the $x$ axis but does not
alter the structure of $|\psi_n^0|^2$. The wavefunctions consist of
two symmetrical contributions from the $A$ and the $B$ sublattices,
and the two parts have a reflection symmetry axis (dashed line). The
position of the reflection symmetry axis of the wavefunctions with
respect to the superlattice potential barriers is crucial in
determining the amplitude of the respective $\kappa_n$ integrals. In
the upper parts of Fig.~\ref{FIG_Broadening} (c), for
$k_y=2m\pi/(6a\sqrt{3}), m=0,1,2,\ldots$ the symmetry axis of
both the $1^{\rm st}$ and the $2^{\rm nd}$ LLs coincides with the
superlattice barrier edge, where the potential changes sign from $+V$
to $-V$. In this case, the contribution to the $\kappa_n$ integral of
the wavefunctions situated to the left and to the right of the
reflection symmetry axis cancel each other and $\kappa_n$ takes on a
minimal value. On the contrary, for $k_y=(2m+1)\pi/(6a\sqrt{3})$ the
reflection symmetry axis coincides with the middle of a superlattice
potential barrier, the wavefunctions from the left and the right sides
of the reflection symmetry axis contribute the same amount when taking
the integral Eq.~\eqref{overlap}, leading to a maximal $\kappa_n$. For
all other $k_y$, the values of $\kappa_n$ fall in between these two
limiting cases. 

The different amplitudes of $\kappa_n$ for different LLs can
be also explained from the lower parts of Fig.~\ref{FIG_Broadening}
(c), by carefully examining the spatial distribution of the
wavefunction with respect to the barrier width. For the $1^{\rm st}$
LL, the contributions left and right of the reflection symmetry axis
extend mainly over a single superlattice potential barrier, in this
case $+V$. 
In the case of the $2^{\rm nd}$ LL, the wavefunction extends over
three barriers. When calculating $\kappa_2$, one subtracts from the
main contribution under the $+V$ barrier the part of the wavefunction
that are extended over the $-V$ barriers. Hence, the amplitude of
$\kappa_1$ is bigger than the amplitude of $\kappa_2$ and,
correspondingly, the broadening of the $1^{\rm st}$ LL is larger than
the broadening of the $2^{\rm nd}$ LL. 

This approach was carried out for several system parameters to check
its validity. We conclude that such an analysis provides an easy way
to find out a-priori, starting only from the unperturbed wavefunctions
of graphene in a magnetic field, which of the LLs will exhibit a large
or a small broadening when a superlattice potential is switched on. 

\begin{figure}[t]
\includegraphics[width=0.93\columnwidth]{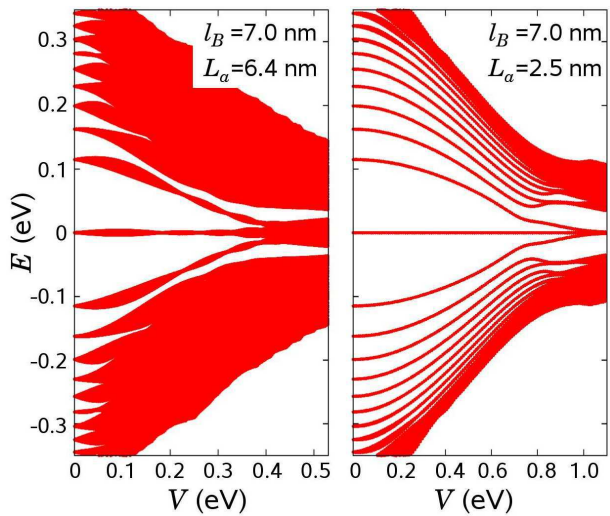}
\caption{The evolution of the low-energy LLs as a function of $V$ for
  graphene superlattice along the zz direction and with $B=13.15$\,T
  ($l_B=7.07$ nm). The superlattice barrier width is $L_a=6.4$\,nm
  (left) and $L_a=2.5$\,nm (right). 
\label{FIG_ZZvsU}}
\end{figure}

When $L_a\approx l_B$, the lowest LLs survive for much higher values
of the superlattice potential strength, as can be seen in
Fig.~\ref{FIG_ZZvsU} (left). Here, $L_a=6.4$ nm (30 zz lines per
barrier) and $l_B=7.07$ nm.
The merging of the $0^{\text{th}}$ and the $\pm 1^{\text{st}}$ LLs
occurs when $V$ takes the value $3U_0/2$ ($=0.42$\,eV for
$L_a=6.4$\,nm), which depends only on the inverse of the superlattice
barrier width according to (\ref{unull}). 
For $V=3U_0/2$ the zero-energy LLs becomes three-fold
degenerate and, as the Fermi energy is scanned from negative to
positive energies, the Hall conductivity presents a step size of $3
(4e^2 /h)$.~\cite{LouieSuperlatB_PRL2009} 

The right side of Fig.~\ref{FIG_ZZvsU}  shows the evolution of the
LLs as $V$ is increased for a system with $L_a<l_B$. The width of the
low energy LLs, although finite, is very small, and depends weakly on
$V$. A general tendency is the bending of the LL energies towards
$E=0$. Again, when $V=3U_0/2$ ($=1.05$\,eV for $L_a=2.5$\,nm, which
are 12 zz lines per barrier), the $0^{\text{th}}$ and the $\pm
1^{\text{st}}$ LLs merge and the zero-energy LL becomes three-fold
degenerate. The small widths of the LLs in the $L_a<l_B$ case can be
explained using a similar analysis of the $\kappa_n$ integrals from
above. In this case, the wavefunctions of the LLs spread over many
superlattice barriers. The contributions of the wavefunctions under
adjacent barriers cancels out when calculating the $\kappa_n$
integrals, which leads to a very small broadening of the LLs. Our
results obtained within the Harper equation method are consistent with
the results from a perturbative approach starting from the continuum
model for graphene.\cite{WKP12} 
There, it was also found that in the case of weak fields (i.e.,
$L_a<l_B$) the matrix elements of the perturbation Hamiltonian do
neither depend on the center nor on the spread of the LL
wavefunctions, which leads to flat bands (i.e., small widths of the
LLs).  

We have performed several calculations for different magnetic field
strengths and superlattice parameters, and the results shown here are
most illustrative. The qualitative behavior of the LL energy spectrum
in the three regimes, namely $L_a>l_B$, $L_a\approx l_B$ and $L_a<l_B$
is robust against changing the system parameters, with the
prerequisite that both the superlattice barrier width and the magnetic
length are much larger than the graphene lattice constant.

\subsection{Superlattice parallel to arm-chair edges}\label{subsection_ACedges}
In the following, we discuss the case of graphene superlattices in a
strong magnetic field with the potential barriers oriented along the
ac direction. When no superlattice is present, the LL energy spectra
for ribbons with zz or ac edges are identical if periodic boundary
conditions are applied. That is because the LLs are a property of the
bulk of the graphene ribbon and not of the edges. However, when a
one-dimensional superlattice potential is switched on, then the system
can be considered to consists of many `sub-ribbons' with internal
edges between them. Although we still investigate superlattice barrier
widths much larger than the graphene lattice constant, we show below
that the orientation of the sub-ribbon edges with respect to the zz or
ac directions of graphene plays an essential role when the magnetic
length $l_B$ is larger than the sublattice barrier width $L_a$. 

\begin{figure}[t]
\includegraphics[width=1.\columnwidth]{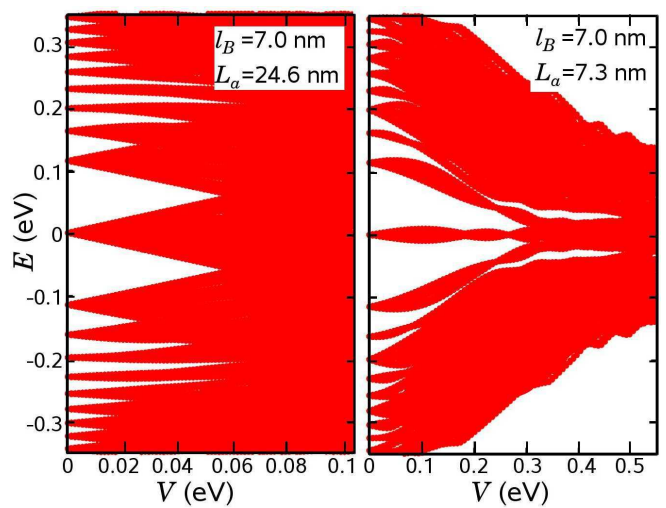}
\caption{The evolution of the low-energy LLs as a function of $V$ for 
  graphene superlattices along the ac direction with $L_a>l_B$ (left)
  and $L_a\approx l_B$ (right). The magnetic flux density is
  $B=13.15$\,T ($l_B=7.07$\,nm) and the barrier widths are
  $L_a=24.6$\,nm corresponding to 200 dimer lines per barrier (left),
  and $L_a=7.3$\,nm corresponding to 60 dimer lines (right).
\label{FIG_EvsVac}}
\end{figure}

Figure~\ref{FIG_EvsVac} shows the energy of the LLs of ac
superlattices as a function of the potential strength for two
different regimes: $L_a>l_B$ (left) and $L_a\approx l_B$ (right). 
Here, $p/q=1/6000$ ($B=13.15$\,T and $l_B=7.07$\,nm), $L_a=24.6$\,nm
corresponds to 200 dimer lines per barrier (left) and $L_a=7.3$\,nm
corresponds to 60 dimer lines per barrier (right). The
behavior of the LLs is qualitatively similar to the case of zz
superlattices: for $L_a>l_B$ the LLs acquire a large broadening
and merge together when increasing the potential strength, and the LL
sequence breaks down for even smaller values of $V$. Also, when
$L_a\approx l_B$, the LLs bend towards zero energy, and their
broadening is not so strong, so that the LL picture survives for
higher values of the superlattice potential strength. However, there 
are some significant quantitative differences between the ac and the
zz cases. For instance, the merging of the $0^{\text{th}}$ and
the $\pm 1^{\text{st}}$ LLs does not occur at  $V=3U_0/2$ any more,
but at values of $V$ which are more complicated to predict from a
continuum model, as they do not depend only on the inverse of the
superlattice barrier width.  

For ac graphene superlattices with strong magnetic field, the
$L_a<l_B$ regime is the most interesting one. In this case, the
electrons travel, in classical terms, over several potential barriers
before closing a cyclotron radius, and the ac edge of each
`sub-ribbon' has an unexpected influence on the energy spectrum of the
LLs. We find a splitting of the $0^{\text{th}}$ LL into two sub-bands
when the potential barrier is
increased. Figure~\ref{FIG_SplittingAC}~(a) illustrates this effect
for an ac superlattice with $L_a=1.47$\,nm (12 dimer lines per
barrier) and $l_B=7.07$\,nm. The $0^{\text{th}}$ LL splits as soon as
$V$ is turned on, and the energy difference between 
the two subbands increases continuously with $V$. For very large $V$,
a splitting of the higher Landau bands was observed too (not shown).
Note also that the higher LLs bend again towards zero energy, and
their broadening is very weak. In the presence of Dirichlet boundary
conditions, no edge states do appear within the energy range between
the split Landau level.

\subsection{Origin of Landau level splitting}
\begin{figure}[t]
\includegraphics[width=1.\columnwidth]{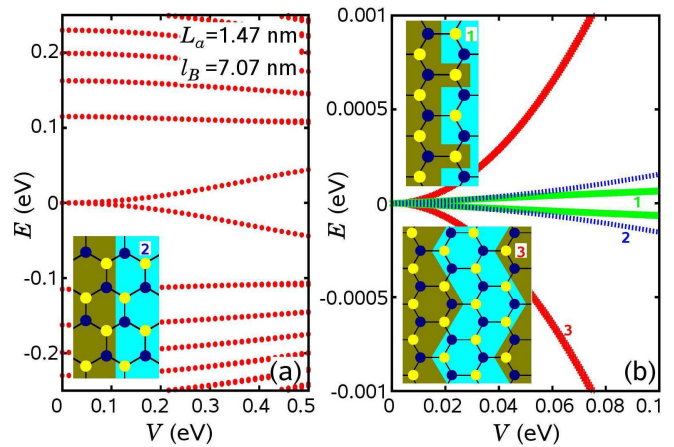}
\caption{(a) Energy of the LLs as a function of the superlattice
  potential strength for graphene ac superlattices with $L_a<l_B$. The
  splitting of the $0^{\text{th}}$ LL is clearly seen. The inset shows
  the barrier edge of the ac-oriented superlattice.  
  (b) The splitting of the  $0^{\text{th}}$ LL of zz superlattices
  with broken symmetry ($L_a^{\rm zz}=1.25$\,nm and $l_B=2.23$\,nm). 
  The smallest splitting (green, labeled 1) corresponds to the 
  system shown in the upper inset, and the largest splitting (red,
  labeled 3) to the lower inset. For comparison, the splitting of an
  ac superlattice with $L_a^{\rm ac}=1.23$\,nm is also shown (blue
  data, labeled 2). The insets in Fig.~\ref{FIG_SplittingAC}~(b)
  schematically show the artificial barrier edges of two zz
  superlattices with broken symmetry that divide the graphene hexagons 
  into 2 and 4 carbon atoms. 
\label{FIG_SplittingAC}}
\end{figure}

A possible explanation for the splitting can be found by carefully
examining the superlattice barrier potential step. For ac
superlattices, the barrier step asymmetrically divides the graphene
hexagons, with 4 atoms on one side and two next-neighbor atoms on the
other side (see the inset of Fig.~\ref{FIG_SplittingAC}~(a)).
In the case of zz superlattices, where the splitting does not occur,
the superlattice barrier symmetrically divides the graphene hexagons,
with 3 carbon atoms under one barrier and the other 3 atoms under the
next barrier with opposite sign. We have considered zz superlattices
with an artificially imposed asymmetry, realized by placing an atom
from each divided hexagon under the adjacent barrier, so that a 4-2
asymmetry is created for all hexagons associated with a barrier
potential step. This configuration is
schematically shown in the upper inset of Fig.~\ref{FIG_SplittingAC}
(b). Now, the energy spectrum clearly shows the splitting of the
$0^{\text{th}}$ LL which is, however, not as strong in magnitude as in
the case of ac superlattices. Another configuration of an artificial
superlattice with the barriers oriented along the zz direction that
divides the step hexagons into 4 atoms on one side and a pair of
next-neighbor atoms on the other side, is shown in the lower inset of
Fig.~\ref{FIG_SplittingAC}~(b). Again, the $0^{\text{th}}$ LL is
split into two sub-bands as the strength of the potential $V$ is
increased, with a splitting magnitude larger compared to the previous
case. 

These examples point already to the origin of this new effect. It is
the absence of inversion symmetry due to the combined influence of
magnetic field and superlattice potential that matters in the ac case.
And in the zz situations discussed above, the artificially imposed
asymmetry that is responsible for the splitting destroys the inversion
symmetry as well.  Once more, a closer look at the wavefunctions in
the absence of the superlattice together with the application of a
(degenerate) perturbation theory treatment up to second order in the
superlattice potential provides a microscopic explanation for the
LL splitting. In both the zz and ac situations, the Landau levels get
broadened in first order perturbation theory due to the superlattice
potential.    
A splitting of the LL, however, is seen in second order only in the ac
case because of the lacking inversion symmetry. The latter is still
present in the zz situation (without an artificial symmetry breaking)
so that energy levels remain degenerate.  
Thus, the ac superlattice induced LL splitting is the generic case,
whereas the particular inversion symmetry present in the perfect
zz orientation turns out to be only a special situation, probably not
met in real samples. We conclude that this LL splitting is a true
lattice effect that was not seen before. The splitting is not to be
found in the usual continuum model descriptions, since there, one
normally cannot distinguish between ac or zz oriented superlattices.

\subsection{Splitting in $3N$ dimer samples independent of $B$}
The splitting for ac superlattices is still present when changing the
system parameters, i.e., $B$, $L_a$, and $V$, as long as we are in the
$L_a<l_B$ regime. Interestingly enough, we find that the splitting of
the  $0^{\text{th}}$ LL is maximal when $L_a$ is chosen such that the
number of ac dimer lines under each superlattice barrier is a multiple
of 3. Moreover, in such cases the energies of the split $0^{\text{th}}$
LL do not change with the magnetic field and depend only on $V$ and
$L_a$. For other values of $L_a$, when the number of ac dimers lines
under each barrier is not a multiple of 3, the splitting is still
present, but the energy difference between  the split sub-bands is one
order of magnitude smaller than in the previous case and depends on
the strength of the magnetic field as $\propto B^2$. 

Experimentally, ideal rectangular superlattices with sharp potential
jumps can hardly be fabricated. To verify that the splitting occurs
also when the barrier edges are smooth, we have performed calculations
considering a width $w$ between two adjacent barriers where the
superlattice potential changes linearly from  $+V$ to $-V$. We find
that when increasing $w$, the energy difference between the split
subbands of the $0^{\text{th}}$ LL decreases, but the splitting is
still present even for $w=L_a/2$.

\subsection{Influence of disorder}
\begin{figure}[t]
\includegraphics[angle=270,width=.95\columnwidth]{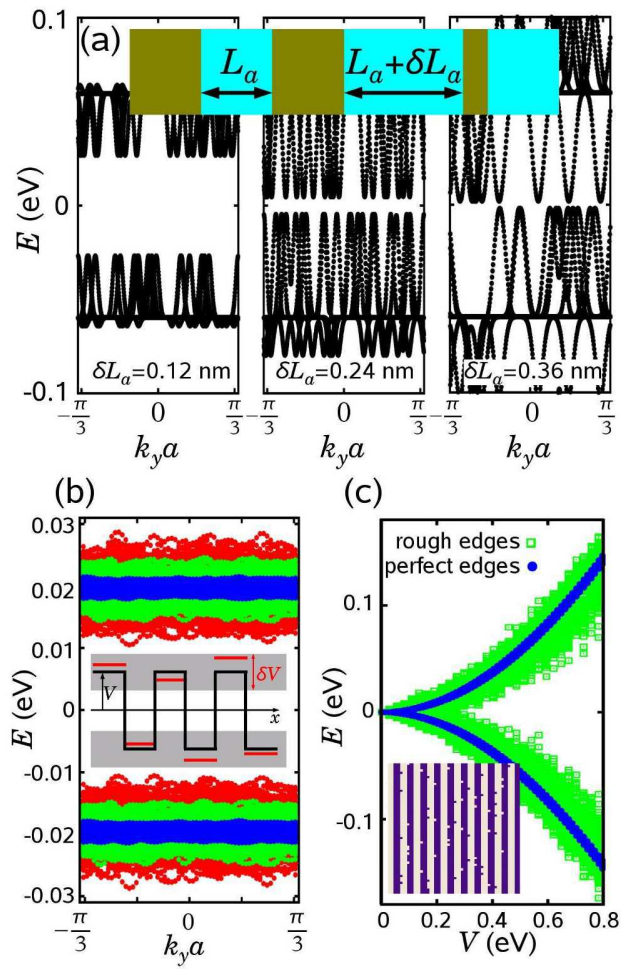}
\caption{(a) The energy spectrum of a graphene ac superlattice with
  barrier-width disorder. For 10\% of the barriers, the width is
  $L_a+\delta L_a$ with $\delta L_a = 0.12$\,nm, 0.24\,nm, and
  0.36\,nm.   
 (b) The energy spectrum of a graphene ac superlattice with
  barrier-height disorder. Here, the potential height varies at
  random between $V-\delta V$ and $V+\delta V$ with $\delta
  V/\text{eV} =0.03, 0.08, 0.13$, and $V=0.3$\,eV. 
 (c) The energy of the $0^{\text{th}}$ LL as function of
  $V$ for ac superlattices with rough edges.
\label{FIG_RandomAll}}
\end{figure}

To make sure that the splitting is not induced by an artificial hidden
symmetry of the Hamiltonian, we have considered different kinds of
superlattice disorder in our calculations. First, we have studied
the case of barrier width disorder, by allowing $10\%$ of the barriers
to have the width $L_a \pm \delta L_a$. The results are shown in
Fig.~\ref{FIG_RandomAll} (a) for a system with $l_B=4.08$\,nm,
$L_a=1.47$\,nm, and $V=0.25$\,eV. Increasing the value of $\delta L_a$
from $0.12$\,nm to $0.36$\,nm results in a broadening of the split
$0^{\text{th}}$ LL. When the value of $\delta L_a$ is further
increased, the broadening overlaps the splitting. The next case of
disorder considered is barrier height disorder. The potential
strengths of the superlattice barriers are allowed to take random
values in an interval $\delta V$ around $\pm V$, as is schematically
shown in the inset of Fig.~\ref{FIG_RandomAll} (b). Again, we find
that this type of disorder does not destroy the splitting, and only 
induces an additional broadening of the sublevels, which grows with
increasing disorder strength. Figure~\ref{FIG_RandomAll} (b) shows
the energy band in the first Brillouin zone of the split
$0^{\text{th}}$ LL for a system with $L_a=1.47$\,nm, $l_B=7.07$\,nm 
and $V=0.3$\,eV. Here, the  disorder strengths is $\delta V=0.03$, $0.08$
and $0.13$\,eV, respectively. Shown are the results for 10 different
disorder realizations. 

Finally, we have considered ac superlattices with rough edges, as
schematically shown in the inset of Fig.~\ref{FIG_RandomAll}~(c). In
this case, the superlattice potential depends on the $y$ coordinate,
and the Harper equation approach cannot be used. Therefore, we have
directly diagonalized the tight-binding Hamiltonian for graphene with
superlattice potential along the ac direction and with 
rough edges. Figure~\ref{FIG_RandomAll} (c) shows the energy of the
split  $0^{\text{th}}$ LL for ac superlattices with rough edges,
compared with the case of perfect edges. In this case, $1.25\%$ of
the atoms along one barrier edge are taken at random and forced to
have a potential strength equal to the one of the next barrier. We
show the results for 100 disorder realizations and conclude that the
splitting is still present in the spectrum. 

The splitting of the $0^{\text{th}}$ LL leads to the prediction of 
a $\sigma_{xy}=0$ plateau to occur in the quantum Hall
conductivity, which may be regarded as a hallmark of ac oriented
graphene superlattices. Most interesting, the splitting occurs only
when $L_a<l_B$. Because $B\sim l_B^{-2}$, this corresponds to magnetic
fields that are strong enough to have LLs, but low enough to be in the
$L_a<l_B$ regime. When increasing $B$, the magnetic length is reduced
and the splitting disappears for $L_a>l_B$. Here, we have presented
the results for $L_a=1.5$ nm, $V$ ranging between 0 and 0.85\,eV, and
$l_B=7.0$\,nm which corresponds to $B=13.1$\,T and are realistic
parameters. Experimentally achievable are $L_a$ between $2-10$\,nm,
and $l_B=5-20$\,nm ($B=2-25$\,T), and superlattice potential strengths
of a few tenths of an electronvolt. Thus it should be possible to
check our findings experimentally. 

\bigskip 

\section{Summary}\label{sectionSummary}
We have investigated the Landau level structure of single layer
graphene in the presence of a one-dimensional electrostatic
square-potential superlattice. The superlattice modulates the Landau
level spectrum in a unique way that is most directly reflected in a
peculiar band broadening, in band bendings, and in an unusual plateau
sequence of the quantum Hall conductivity. We have shown that,
depending on the magnitude of the superlattice barrier width $L_a$
with respect to the magnetic length $l_B$, the energy band structure
of the LLs changes dramatically. In general, when $L_a>l_B$, the LLs
are quickly broadened and merge together for small values of the
superlattice potential strengths. In the opposite regime, when
$L_a<l_B$, the LLs survive for much higher values of $V$, with a
general tendency of the higher-energy LLs to bend towards zero energy. 

The orientation of the superlattice barriers with respect to the zz or  
ac directions of graphene also plays a crucial role when $L_a<l_B$. 
That is because the superlattice divides the system into many
sub-ribbons of width $L_a$ that become decoupled when increasing the 
strength of the superlattice potential $V$. When the magnetic length
is larger than the sub-ribbon width, an electron travels over many
barriers before it closes a cyclotron radius. Therefore, the results
differ depending on the sub-ribbon edge type the electron
encounters at each barrier jump. When the sub-ribbons have ac edges,
we found a novel effect that originates from the interplay between
graphene's hexagonal lattice and the additional superlattice potential
barriers and can, therefore, not be found in the usual Dirac-fermion
continuum model description. The new observation is the splitting of
the zeroth LL which occurs with increasing the superlattice potential
strength. Alternatively, one can tune the magnetic flux density
instead and keep the superlattice potential strength fixed. This
intriguing effect is linked to the absence of inversion symmetry in
the ac case due to the presence of both a superlattice potential and
the magnetic field. The parameters that we used here are
experimentally accessible, and the peculiar features of the electronic
structure may be tested directly in transport or optical experiments.  

Finally we mention that according to further calculations (results not
shown), a splitting of the zeroth Landau level occurs also in the
presence of truly two-dimensional (chess-board type) superlattices,
which remains robust even in the presence of additional on-site
disorder. This observation opens the route to consider the charge 
density fluctuations\cite{Mea08,Dea11} occurring in real graphene
samples on a length scale between $20$\,nm and $30$\,nm as a natural
2D-superlattice that replaces the artificial one investigated in our
model calculations. We then suggest that if the magnetic field is
tuned such that the size of the charge `puddles' and the magnetic
length had the required ratio, the resulting gap opening could be
responsible for an insulating behavior and the diverging resistance at
the Dirac point with the accompanying $\sigma_{xy}=0$ Hall plateau as
has been observed previously in experiments by Checkelsky \textit{et
  al.}\cite{CLO08,CLO09}


\end{document}